\documentstyle[aas2pp4]{article}

\slugcomment{Submitted to ApJ, Part I}

\lefthead{Gibson \& Mould}
\righthead{Nucleosynthesis \& the Galactic Halo}

\begin{document}

\title{The Chemical Residue of a}

\vspace{-7.0mm}
\title{White Dwarf-Dominated Galactic Halo}

\author{Brad K. Gibson \& Jeremy R. Mould}
\affil{Mount Stromlo \& Siding Spring Observatories, \break 
Institute of Advanced Studies, \break
Australian National University, \break
Weston Creek P.O., Weston, ACT, Australia  2611}

\vspace{2.0mm}
\centerline{Accepted for publication in ApJ, Part I}

\def\spose#1{\hbox to 0pt{#1\hss}}
\def\simlt{\mathrel{\spose{\lower 3pt\hbox{$\mathchar"218$}}
     \raise 2.0pt\hbox{$\mathchar"13C$}}}
\def\simgt{\mathrel{\spose{\lower 3pt\hbox{$\mathchar"218$}}
     \raise 2.0pt\hbox{$\mathchar"13E$}}}
\def\eg{{\rm e.g. }}
\def\ie{{\rm i.e. }}
\def\etal{{\rm et~al. }}

\begin{abstract}
Halo initial mass functions (IMFs), heavily-biased toward white dwarf (WD)
precursors (\ie $\sim 1\rightarrow 8$ M$_\odot$), have been 
suggested as a suitable mechanism for explaining microlensing statistics 
along the line of sight to the LMC.  Such IMFs can apparently be invoked
without violating the observed present-day WD luminosity function.  
By employing a simple chemical evolution argument, we demonstrate that
reconciling the observed halo Population II dwarf abundances (\ie
[C,N/O]$\approx -0.5$), with that expected from the postulated ``WD-heavy'' IMF
(\ie [C,N/O]$\simgt +0.5$), is difficult.
\end{abstract}


\section{Introduction}
\label{introduction}

Analysis of the first year's MACHO (Alcock \etal 1993) data led some
to conclude that low-mass stars and brown dwarfs could be responsible for the
microlensing events seen along the line-of-sight to the Large Magellanic Cloud
(\eg Fujimoto \etal 1995; M\'era, Chabrier \& Schaeffer 1996).  
Reconsideration of this position seems apparent in light of the release of the 
second year's worth of MACHO data (Alcock \etal 1997)
which points to substantially higher-mass ``lenses''.

A dark baryonic halo comprised primarily of white dwarfs (WDs) has been
considered in this context on more than one occasion in the past.  
Adopting a variant of Larson's (1986) Galactic,
bimodal, initial mass function (IMF),
Hegyi \& Olive (1986) clearly demonstrated that such a scenario was untenable,
for IMFs with lower-mass cutoffs of 2 M$_\odot$, based
simply upon an overall overproduction of heavy elements.  Unlike Hegyi \& Olive
(1986), who adopted an upper-mass limit of 100 M$_\odot$, 
Ryu \etal (1990) considered truncating this limit at progressively smaller
values, until metal overproduction was minimized, concluding finally that only
very specific, and limited, ranges were allowed.  
Further arguments against WD-dominated halos came from
Charlot \& Silk's (1995) examination of their high-redshift photometric
signatures.  Charlot \& Silk found that halos whose WD mass fraction was
$\simgt 10$\% would clearly violate the galaxy number counts.

Recently, Adams \& Laughlin (1996), Chabrier, Segretain \& M\'era (1996), 
and Fields, Mathews \& Schramm (1996), 
have explored the ramifications of a WD precursor-dominated halo
IMF, ensuring that each of their respective favored models did not lead to 
inconsistencies with the observed present-day 
halo WD luminosity function.  The \it detailed \rm
nucleosynthesis implications were
beyond the scope of these initial studies 
(\eg global metallicity Z was cursorily
considered, but the evolution of specific elements was not).
Our follow-up work, described herein,
is a first step in redressing this omission; it
is not meant to be exhaustive, but does serve to indicate potential problems
with the WD precursor-dominated IMF scenario, not fully appreciated in these
analyses.

In Section \ref{analysis}, we briefly describe the chemical evolution code
adopted.  We then concentrate on the early interstellar medium (ISM) evolution
of carbon, nitrogen, and oxygen, contrasting the implied behavior as a
function of adopted IMF with observations of metal-poor halo dwarfs.  Finally,
qualitative arguments based purely upon the implied mass of ejected gas from
the WD-precursor dominated IMF is presented.
Our results are summarized in Section \ref{summary}.

\section{Analysis}
\label{analysis}

\subsection{Background}
\label{background}

We adopt Gibson's (1996a,b) chemical evolution package, in order to follow
the temporal history of CNO abundances in our simple Galactic halo model.  
The star
formation rate is presumed to be proportional to the available gas mass, with a
constant of proportionality (\ie the astration parameter) $\nu=10$ Gyr$^{-1}$.
Such a formalism corresponds to an exponential star formation rate of the form
$\psi\propto e^{-t/\tau}$, with $\tau\approx 0.11$ Gyr for $t\simlt 0.4$
Gyr, and $\tau\approx 2.56$ Gyr for $t\simgt 0.4$ Gyr.  Parallel calculations
were made with higher and lower values for $\nu$ and $\tau$, to ensure that our
conclusions were not dependent upon these template values (which they were
not).

We have not considered the role played by Type Ia supernovae (SNe)
in what follows,
as we will be primarily concerned with the CNO abundances, none of which are
supplied by Type Ia's in any important quantity.  Where this could be important
though would be in the calculation of the present-day Type Ia SNe rate (and its
accompanying increase in the mass of iron made available for subsequent
generations of star formation).  For example, the
favored Chabrier \etal (1996) IMF discussed explicitly in the following
subsection has a factor of two more mass tied up in the
$3\rightarrow 16$ M$_\odot$ regime, a range generally accepted as the progenitor
mass range for most Type Ia binary-progenitors, regardless of whether mass
transfer or WD-merging is the dominant mechanism (Greggio \& Renzini
1983; Tornamb\`e 1989).

The key ingredients in our modeling, as shall be elucidated upon in the
following subsections, will be the adopted IMF and stellar yields.  Before
commenting upon their significance, let us first list briefly the fundamental
observational constraints that we shall be concerned with in this paper.
The observational datasets collated by
Timmes \etal (1995) show that
[C/O] in the halo dwarfs ranges from -1.2 to +0.3 dex for [Fe/H]$\simlt$-1.5
\footnote{On the other hand, Gratton \& Caretta (1996) claim that the
range of [C/O] in halo dwarfs is actually considerably smaller than this -- \ie
[C/O]$\approx -0.6\rightarrow -0.2$, for [Fe/H] down to $\sim -2.0$ -- a fact
which lends even further credence to the conclusions which follow.};
[N/O], in the same metallicity regime, goes from -1.7 to +0.7 dex.
\footnote{The halo dwarf [C,N/O] ratios were estimated from Timmes et~al.'s
(1995) Figures 11, 13, and 14, using the relation 
[C,N/O]$\equiv$[C,N/Fe]-[O/Fe].}
\it The mean observed value for both ratios
is [C/O]$\approx$[N/O]$\approx -0.5$, with the majority ($\sim 80$\%)
of dwarfs lying within
$\pm 0.3$ dex of the mean.  \rm
It is important to echo Timmes et~al.'s remarks, and note
that field and halo giants are simply not reliable indicators of any \it ab
initio \rm abundance pattern, as mixing processes along the giant branch
can dramatically alter both carbon and nitrogen.

\subsection{Initial Mass Functions}
\label{IMFs}

\begin{figure}[ht]
\epsscale{1.0}
\plotone{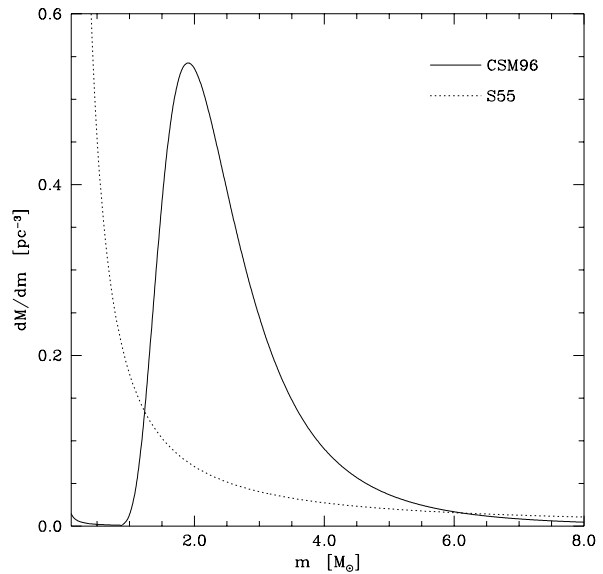}
\caption[fig1.eps]{
Comparison of the canonical Salpeter (1955) IMF (dotted
curve), normalized to unity over the range $0.1\le m\le 40.0$ M$\odot$,
with the WD precursor-dominated IMF proposed by Chabrier \etal (1996)
(solid curve).
\label{fig:fig1}}
\end{figure}

For brevity, we shall restrict our analysis to two different IMF forms --
-- the aforementioned favored WD
precursor-dominated IMF of Chabrier \etal (1996),\footnote{We could just as
easily have chosen to use the Adams \& Laughlin (1996) or Fields \etal (1996)
IMF formalism, 
but their similarity to that of Chabrier et~al.'s (1996) means that our
conclusions are \it not \rm dependent upon this choice.}
which in turn will be contrasted against
that of the canonical Salpeter (1955) IMF.  Both are illustrated in Figure
\ref{fig:fig1}, normalized to unity over the mass range $0.1\rightarrow 40.0$
M$_\odot$, clearly demonstrating, better than any words can, exactly
how these two IMFs differ.  It is apparent that Chabrier et~al.'s IMF
has effectively no component of sub-solar mass stars, while Salpeter's has
almost 2/3 of the mass locked-up below 1 M$_\odot$.  At the high mass end,
Salpeter's IMF has $\sim 25\times$ as much mass locked into Type II SNe
progenitors (\ie $m\simgt 11$ M$_\odot$).

Not surprisingly, the Chabrier \etal (1996) IMF, because of its resultant
dominance by WDs\footnote{$\sim 50$\% of the present-day dynamical
mass of the halo ($M_h\approx 10^{12}$ M$_\odot$ -- Freeman 1996) is assumed to
be locked into WDs, for the Chabrier \etal (1996) IMF under
consideration here.}, is far more successful at replicating the inferred
present-day
Galactic halo mass-to-light ratio $M/L\simgt 100$ (\eg Freeman 1996, and
references therein), in the absence of a large non-baryonic component,
than that of
Salpeter's (1955).  Adopting the isochrones of Bertelli \etal (1994), and
following the photometric evolution prescription of Gibson (1996a), we found
that for the models to be discussed later in this section (and Figure
\ref{fig:fig4})
$M_h/L_{\rm V}\approx 300$ (Chabrier \etal IMF) and $M_h/L_{\rm V}\approx 6$
(Salpeter IMF).

\subsection{Stellar Yields}
\label{yields}

Besides the IMF selection, the other key ingredient to
our modeling is the adopted nucleosynthetic yields.  We have chosen Woosley \&
Weaver's (1995) metallicity-dependent yields for Type II SNe ejecta,\footnote{
The published Woosley \& Weaver (1995) Type II
SNe models do not produce any primary nitrogen.
Nitrogen yields, though, can be a strong function of the
arbitrary prescription chosen for convective overshooting (Arnett 1996);
Models kindly
provided by F. Timmes, based upon the same code used by Woosley \& Weaver
(1995) but with an
enhanced overshooting prescription (although one not at odds with
observation), show that primary nitrogen was produced in their models with
$m\ge 30$ M$_\odot$, but \it only \rm for metallicities Z$\simlt 0.01$.  This
massive star component of primary nitrogen has been used in lieu of the
published values.
}
although
this is a relatively inconsequential decision as Type II SNe play a fairly
unimportant role when coupled with the Chabrier \etal (1996) IMF.  For the
lower-mass asymptotic giant branch (AGB) 
precursors, our default yield
prescription is that due to van den Hoek \& Groenewegen
(1996).  We have considered competing prescriptions (\ie Marigo \etal 1996 and
Renzini \& Voli 1981), a point to which we return to briefly, below, but
stress that our conclusions are not dependent upon the AGB yields selected.

\begin{figure}[ht]
\epsscale{1.0}
\plotone{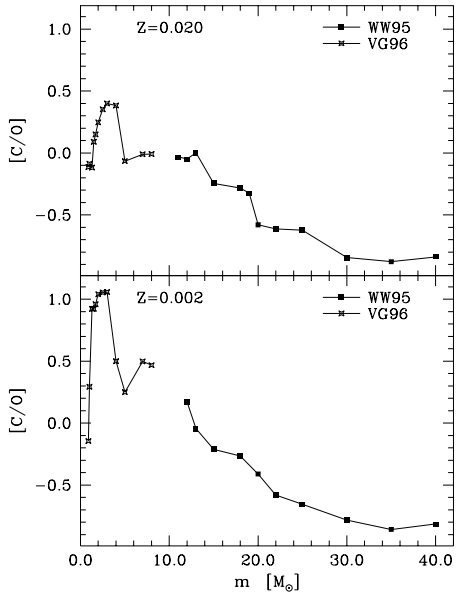}
\caption[fig2.eps]{
Ratio of ejecta carbon to oxygen predicted by 
the Type II SNe models of Woosley \&
Weaver's (1995) ([Fe/H]=-1.0, $m\simgt 10$ M$_\odot$),
contrasted with low- and intermediate-mass AGB models of van den Hoek \&
Groenewegen (1996).  Two different metallicities (solar -- upper panel; 1/10
solar -- lower panel) are highlighted.  Recall that the halo dwarfs
show the abundance pattern [C/O]$\approx$[N/O]$\approx$-0.5, albeit with
large scatter.
\label{fig:fig2}}
\end{figure}

\begin{figure}[ht]
\epsscale{1.0}
\plotone{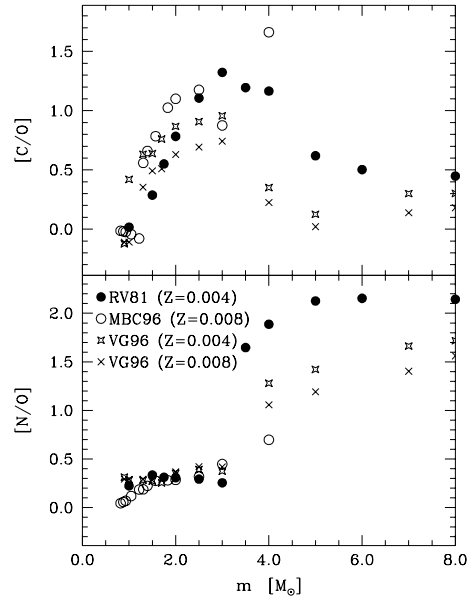}
\caption[fig3.eps]{
Ratios of ejecta carbon and nitrogen 
to oxygen predicted by the sub-solar metallicity AGB models of
Renzini \& Voli (1981) and Marigo \etal (1996).  Recall that the halo dwarfs
show the abundance pattern [C/O]$\approx$[N/O]$\approx$-0.5, albeit with
large scatter.
\label{fig:fig3}}
\end{figure}

Figure \ref{fig:fig2} illustrates the stellar ejecta's
carbon-to-oxygen ratio, as a function of initial mass $m$, for stars of solar
metallicity (top panel), and one tenth solar metallicity (bottom
panel).  Even a cursory
inspection of Figure \ref{fig:fig2} allows one to anticipate
problems that will arise in reconciling \it any \rm intermediate-mass-biased 
IMF, with the average halo dwarf observations of [C,N/O]$\approx -0.5$
-- specifically, the \it only \rm portion of the IMF which could possibly leave
the imprint of such a carbon underabundance relative to oxygen in the halo
dwarfs comes from $m\simgt 15$ M$_\odot$.  In particular, any IMF which is
predicated upon being heavily biased toward WD precursors (\ie
$m\approx 1\rightarrow 5$ M$_\odot$) cannot help but \it overproduce \rm carbon
with respect to oxygen.  This is demonstrated even more so in Figure
\ref{fig:fig3} which shows a close-up of the mass region in question.

Figure \ref{fig:fig3} shows examples of
the low metallicity AGB yield predictions of van den Hoek \& Groenewegen
(1996), Marigo \etal (1996), and Renzini \& Voli (1981), 
for both [C/O] and [N/O].
The first two references detail the differences in the
models, but again though, for our purposes, whether we choose one
compilation over the other in no way leads to sub-solar abundance ratios of
carbon or nitrogen with respect to oxygen, regardless of how one arbitrarily
distributes mass in the $m\simlt 8$ M$_\odot$ regime.

An immediate caveat springs to mind at this point -- 
our conclusions rest squarely upon the applicability of the relevant
yield compilations.  
If the low-mass stellar evolution models upon which these compilations are
based could be shown to be severely in
error, then one could conceivably relax the argument of the previous paragraph.
In this vein, Fujimoto \etal (1995) rightly note that the evolution
of Z=0 intermediate-mass stars may be quite different from simply
extrapolating the tabulated solar and mildly sub-solar metallicity models 
to arbitrarily low Z.\footnote{Specifically, we are forced to 
extrapolate the tabulated yields to metallicities lower than [Fe/H]=-1.00,
-0.40, and -0.70, for van den Hoek \& Grownewegen (1996), Marigo \etal (1996),
and Renzini \& Voli (1981), respectively, to arbitrarily low halo 
metallicities.  These values represent the minimum
[Fe/H] considered in the respective models.}
On the other hand, Fujimoto \etal
claim that \it extreme \rm nitrogen-rich carbon stars
should be the outcome of primordial composition evolution, which referring to
the bottom panel of Figure \ref{fig:fig3} will only drive the expected 
[N/O] ejecta
from this low-metallicity WD precursor-dominated IMF to values even further
from the observed halo dwarf values of [N/O]$\approx -0.5$.
Regardless, this entire extrapolation procedure is, at some level, a leap of
faith, and we reserve the right to modify our conclusions once primordial
metallicity AGB nucleosynthetic yields become available!

\subsection{Results}
\label{results}

\begin{figure}[ht]
\epsscale{1.0}
\plotone{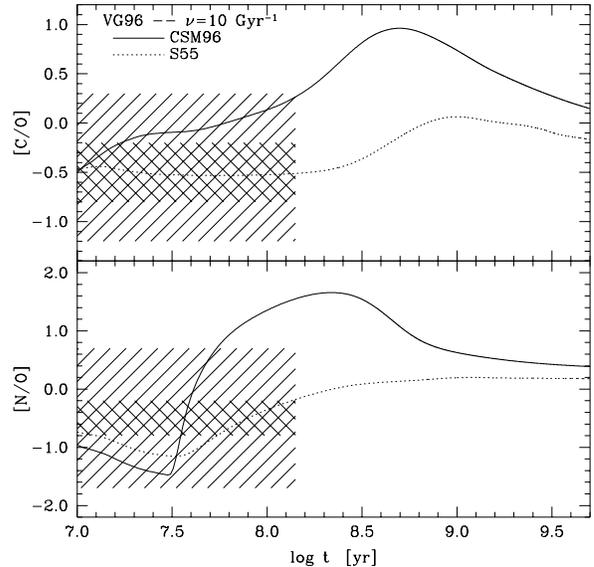}
\caption[fig4.eps]{
Evolution of the ISM [C/O] and [N/O] for the model
described in Section 2.  The two IMFs of Figure 1 are shown -- Salpeter (dotted
curve) and Chabrier \etal (solid curve).  The Renzini \& Voli (1981) yields for
low- and intermediate-mass stars were assumed.  The observational constraints,
from Timmes \etal (1995), are indicated by the shaded regions.
See text for details.
\label{fig:fig4}}
\end{figure}

The [C,N/O] evolution of our model halo ISM, under the input parameters
outlined above, is illustrated in Figure \ref{fig:fig4}.  The solid curve
represents the expected behavior utilizing the favored Chabrier \etal (1996)
IMF, whereas the dotted curve is the corresponding result when implementing the
Salpeter (1955) IMF.  The observational constraints, from the compilation of
Timmes \etal (1995), are indicated by the shaded regions.  All of the $\sim
150$ halo dwarfs in Figures 13 and 14 of Timmes \etal lie within the bounds of
the outer shaded region; $\sim 80$\% of the sample lies within
the inner region.  The ISM metallicity at $\log t\approx 8.15$ is
Z$\approx$0.001; only a small halo stellar component exists at
metallicities higher than this (Fields \etal 1996), which is why we have ended
the shaded regions there.

A few general comments regarding the morphological behavior of the curves in
Figure \ref{fig:fig4} can be made now -- 
the turnoff-time for an 8 M$_\odot$ star is approximately $\log t=7.56$
(Schaller \etal 1992), which corresponds to the point in the bottom panel at
which [N/O] begins its initial dramatic increase, not surprisingly.
The parallel increase in [C/O] is delayed somewhat relative to [N/O], until 
$\sim 4$ M$_\odot$ stars start turning off the main sequence 
(\ie $\log t\approx 8.16$).  
Again, this could have been anticipated by referring
back to the carbon and nitrogen behavior of 
Figure \ref{fig:fig3}.  The decline beyond $t\approx 0.3$ Gyr
coincides with the expected dilution in [C,N/O] as stars with masses below
$\sim 2$ M$_\odot$ start returning their ejecta to the ISM (recall Figure
\ref{fig:fig3}).
The ISM [Fe/H] in Figure \ref{fig:fig4} attains the values
-2.3, -2.0, -1.0, and +0.0, at $\log t=$7.78, 7.89, 8.27, and 9.31,
respectively.  

\begin{figure}[ht]
\epsscale{1.0}
\plotone{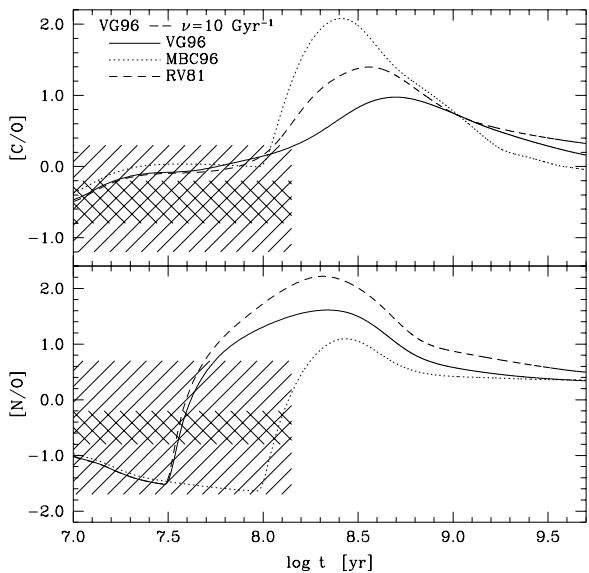}
\caption[fig5.eps]{
Evolution of the ISM [C/O] and [N/O] for the model
described in Section 2.  Three different low and intermediate-mass stellar
yields are compared -- Renzini \& Voli (1981), Marigo \etal (1996), and van den
Hoek \& Groenewegen (1996).
The Chabrier \etal (1996) IMF is adopted in all cases.
The observational constraints,
from Timmes \etal (1995), are indicated by the shaded regions.
See text for details.
\label{fig:fig5}}
\end{figure}

Figure \ref{fig:fig5} parallels that of Figure \ref{fig:fig4}, but now shows
how the chemical evolution depends upon the choice of AGB yields.  The
behavior in each case is qualitatively similar to that described in the previous
paragraph, and indeed, could have been anticipated from Figure \ref{fig:fig3}
-- (i) [C/O] peaks several tenths in dex higher with the Marigo \etal (1996)
yields, because of the very high [C/O] in their [Fe/H]=-0.40, $m=4$ M$_\odot$
model.  van den Hoek \& Groenewegen's (1996) carbon and nitrogen yields are
mildly reduced, 
in comparison with those of Renzini \& Voli (1981) (see Figure 
\ref{fig:fig3}, which results in their somewhat lower [C/O] in Figure
\ref{fig:fig5}.
(ii) The [N/O] behavior, when using Marigo et~al., is improved over that of the
``competitors'' (although, recall that its [C/O] behavior was worse),
as inspection of the lower
panel of Figure \ref{fig:fig3} intimates; Marigo \etal have no [N/O]$\simgt
+1$ models for $m\simgt 3.5$ M$_\odot$ (enhanced convective overshooting being
the primary cause).
Such a difference leads to [N/O] ranging from $\sim +0.0$ to $\sim +1.0$, when
using Marigo et~al., as opposed to the $\sim +0.0$ to $\sim +2.0$ we see in
Figures \ref{fig:fig4} and \ref{fig:fig5}, when using Renzini \& Voli or van
den Hoek \& Groenewegen.  
We do not show the Salpeter (1955) curves in Figure \ref{fig:fig5} as they are
very similar to that shown in Figure \ref{fig:fig4}; the Salpeter IMF
is relatively
insensitive to the AGB yield selection.
When using the van den Hoek \& Groenewegen AGB models,
all yields for times $\log t\simlt 8.15$ were based upon extrapolating beyond
their minimum Z model (\ie Z=0.001); for the Marigo \etal models, the
extrapolation ``regime'' was $\log t\simlt 8.35$, and for Renzini \& Voli,
$\log t\simlt 8.24$.

Recalling the observational constraint that Population II halo dwarfs have
an intrinsic mean [C,N/O]$\approx -0.5$, the conclusion to be
inferred from the Chabrier \etal (1996) curves of Figure \ref{fig:fig4} is
fairly obvious.  \it The combination of the WD precursor-dominated IMF, with
the AGB yields of van den Hoek \& Groenewegn (1996),
leads to inevitable overproduction of
carbon and nitrogen, relative to oxygen, by factors of $\sim 5\rightarrow
30$ and $\sim 8\rightarrow 60$, respectively, 
for all times $t\simgt 0.1$ Gyr. \rm  We stress that the van den Hoek \&
Groenewegen yields are the most favorable compilation in this regard;
adopting the Renzini \& Voli or Marigo \etal yields only exacerbates the
problem (by a \it further \rm factor of $\simlt 10$), 
especially the former, as far as nitrogen goes, and the latter, as far
as carbon goes.  On very short timescales, then,
this enrichment of the halo ISM should be reflected in the present-day
Population II halo dwarfs, \footnote{There is a class of dwarfs in the 
halo (De Kool \& Green 1995) with C$>>$O;
Dearborn \etal (1984) speculated that the prototype of these dwarf carbon stars
might bear the nucleosynthetic imprint of Population III-like pollution.
The origin of these stars is now,
however, generally attributed to binary mass-transfer from an AGB
primary (Green \& Margon 1994).}
which, as was noted in Section \ref{background}, is simply not the case. 
The Salpeter (1955) IMF curves of Figure \ref{fig:fig4}
are not meant to be adjudged to represent
the true halo IMF, but are merely included as a comparison, indicating that any
form of Population III-style ``pre-enrichment'', if it even exists, is 
almost certainly based upon something resembling this more conventional form,
as opposed to Chabrier et~al.'s.

While we have not shown it graphically, 
extending the timescale for star formation, by decreasing $\nu$
by a factor of ten, say, only increases
an already problematic discrepancy between
observation (\ie [C,N/O]$\approx -0.5$) and theory (\ie [C,N/O]$\simgt +0.5$)
by lengthening the [C,N/O]$\simgt +0.5$ phase by a factor of $\sim 3$ over that
shown in Figures \ref{fig:fig4} and \ref{fig:fig5}.

Finally, we have only been concerned with recovering the
mean halo dwarf abundance 
[C,N/O]$\approx -0.5$; we should remind the reader that there is a
fairly wide spread in abundance ratios at these low metallicities 
(\ie [Fe/H]$\simlt -1.5$), with
[C/O] as low as -1.2 dex being encountered (recall, though, our first
footnote).  This is a factor of $\sim 2$ lower
than \it any \rm
single model in Woosley \& Weaver's (1995) grid; there is \it no \rm
possibility of accounting for this tail of the population with their models.
The situation would become
considerably worse if we were to adopt the Langer \&
Henkel (1995) Type II SNe yields, as they are consistently $\sim 5\rightarrow
8\times$ greater in [C/O] than Woosley \& Weaver (1995), for $m\simgt 20$
M$_\odot$.  Maeder's (1992) [C/O] is typically 50\% greater than Woosley \&
Weaver's (1995), for $m\simlt 40$ M$_\odot$.  All of the above may be pointing
to some underlying deficiencies in the existing stellar models, or perhaps
postulating pollution from very massive stars (\ie $m\simgt 40\rightarrow 100$
M$_\odot$ -- a mass regime
to which Woosley \& Weaver's (1995) grid does not apply) 
will be necessary.

\subsection{Hiding the Pollution}
\label{escape}

\subsubsection{Absorbing the Residue}
\label{absorption}

The above constraint on the WD-enriched luminosity function
could \it possibly \rm be relaxed for a very
rapid collapse variant of the classical Eggen, Lynden-Bell \& Sandage
(1962) halo formation model. If star formation in the halo is complete
within $\simlt 70$ Myrs (see Figure 4), the C,N products of the
WD-enriched luminosity function could simply fail to be incorporated into
the halo population by virtue of the longer evolutionary timescales of
intermediate mass stars. These C,N products are inevitable, of course,
but if they are lost to the halo and incorporated into the disk ISM,
they might be diluted away by the primordial composition of the proto-disk.
Unfortunately, this ``escape'' clause would appear to have at least one
major problem, independent of any abundance arguments.

If we were to arbitrarily halt 
star formation at $\sim 70$ Myrs, for the Chabrier
\etal (1996) IMF model of Figure \ref{fig:fig4}, one could argue that the
abundance constraints (\ie [C/O]$\simlt +0.3$ and [N/O]$\simlt +0.7$) were not
violated excessively, and, as it turns out, that the mass
of the halo tied up in remnants (primarily WDs, with total mass 
$\sim 2\times 10^{11}$ M$_\odot$) was not
inconsistent with the microlensing statistics.  
Where this scenario suffers is in the sheer mass of gas postulated to
settle to the disk.
Ignoring any non-baryonic component to the halo, the
model of Figure \ref{fig:fig4} requires an initial gas mass of
$\sim 10^{12}$ M$_\odot$, in order to build up this halo WD mass of $\sim
2\times 10^{11}$ M$_\odot$, when star formation is assumed to halt at $t\approx
70$ Myrs.\footnote{Reducing the star formation efficiency $\nu$
by a factor of ten,
say, does not help.  The $t_{\rm max}\approx 70$ Myrs was chosen to avoid
overproducing C and N; this holds, roughly, regardless of the value of $\nu$.
Unfortunately, because of the factor of ten lower star formation rate, we end
up with approximately a factor of ten lower mass tied up in WDs after $\sim 70$
Myrs.}
The resultant halo stellar (\ie non-WD) mass is $\sim 10^9$
M$_\odot$, in agreement with that observed (Freeman 1996); this still leaves
$\sim 8\times 10^{11}$ M$_\odot$ of gas to absorb!  Bearing in mind
that the present-day mass of the thin+thick disk is only $\sim 0.6\times
10^{11}$ M$_\odot$, it should be readily apparent that 
such halo-to-disk gas ``absorption'' scenarios, at least of this magnitude, are
not a viable option.

\subsubsection{Banishing the Residue}
\label{emission}

Halo \it outflows\rm,
similar to those expected during
the early evolution of ellipticals (\eg Gibson 1996a,b, and references 
therein), would appear to be
a viable alternative to the disk ``incorporation'' of 
Section \ref{absorption}.  
Fields \etal (1996) have recently presented just such an hypothesis.
A detailed accounting of their work is beyond the scope of
this paper, but we do wish to draw attention to two potential problems:

(i) Fields \etal adopt the instantaneous recycling approximation; 
by assuming that
the ejecta from AGB stars is returned on the same timescale as that from
Type II SNe, they overestimate the local gas mass available for heating
(and outflow) from the nearby SNe.  In reality, the timescales are an order
of magnitude different, which means that the bulk of the AGB ejecta (which
itself is the bulk of the gas being returned, for the IMF in question) never
experiences the local SN heating, and it would seem
unlikely that planetary nebulae ejection could provide the necessary
energy (Van Buren 1985).  

(ii) More importantly, 
Fields \etal adopt Maeder's (1992) yields, assuming
a black hole cut-off of $m=18$ M$_\odot$, thereby
avoiding the enrichment from precursors above this level (for halo
metallicities in this mass range, stellar winds prior to the core collapse do
not enrich the ISM in heavy elements 
-- Maeder 1992).  The minimum predicted halo
dwarf [C/O] can only be as low as the minimum individual contributing star's
[C/O] yield.  Because Fields \etal are restricted to Maeder's (1992) low
metallicity $m=9\rightarrow 18$ M$_\odot$ models, which span
[C/O]$\approx +1.3\rightarrow -0.2$, 
the minimum [C/O] possible within their framework is $\sim -0.2$.
In reality, because their IMF puts far more weight on
the 9 M$_\odot$-end, as opposed to the 18 M$_\odot$-end, they will inevitably
predict extremely carbon-rich [C/O] ratios ($>>+0.5$) for the halo.  Fields
\etal
appear to have neglected the CNO-enrichment from AGB stars in their 
code, but we
need only refer back to Figure \ref{fig:fig3} to see that rectifying
this omission will not aid in lowering the predicted [C/O] values.
The Fields \etal outflow model \it may \rm provide a means
for hiding the bulk of the halo gas (although see point (i) above), and it may
indeed be consistent with the distribution of global metallicity ``Z'' in the
halo, but it \it must \rm be inconsistent with the
[C,N/O]$\approx -0.5$ constraint from the halo dwarfs, in much the same way
that our models of Figure \ref{fig:fig4} and \ref{fig:fig5} were.

In conclusion, it is difficult to envision a simple
scenario which would
allow one to (i) create $(\sim 2\rightarrow 5)\times 10^{11}$
M$_\odot$ of halo WDs (as favoured by Adams \& Laughlin 1996, Chabrier \etal
1996, and Fields \etal 1996); (ii) do so on a very
short timescale (to avoid the abundance ratio problems); and
(iii) not produce more than a few times $10^{10}$ M$_\odot$ of C,N-enriched
unincorporated ejecta -- not to mention the residual unincorporated 
primordial composition halo gas -- to avoid violating the Galactic disk
mass-constraint.
The combination of points (i) and (ii) always results in approximately an
order-of-magnitude overproduction of ``hidable'' gas.\footnote{By process of
elimination, if one were determined to retain the notion of a purely
baryonic dark
halo, one might be tempted to side with De Paolis \etal (1997, and 
references therein), and throw support behind cold molecular clouds as the
``hiding'' place for the bulk of the halo's dark matter.}

\section{Summary}
\label{summary}

As a guide to future studies of chemical evolution of the Galactic halo, we
note that a
WD-precursor dominated IMF leads to an inevitable pollution of the
halo ISM, at the levels [C,N/O]$\approx +0.0\rightarrow +1.5$, in timescales
$t\simlt 0.1$ Gyr.  \it If \rm we interpret this IMF as Population 
III-related, one can construct models which are 
in agreement with the present-day WD luminosity
function and the microlensing statistics, as stressed by Chabrier \etal (1996),
and indeed with the inferred present-day halo mass-to-light ratio, as noted in
Section \ref{IMFs}.  On the other hand, 
reconciling the implied
nucleosynthesis with the observed [C,N/O] abundance pattern in the Population
II halo dwarfs appears untenable.
Invoking the argument that the above scenario could be retained, provided the
halo star formation timescale was exceedingly short ($\simlt 70$ Myrs) and
that any subsequent
C,N-enriched gas diffused to the disk, thereby avoiding being locked
into any of the Population II halo dwarfs, fails on the
grounds that the sheer mass of gas that would need ``hiding'' in the disk
would be up to an order of magnitude more massive than the present-day disk
itself.  Scenarios whereby this enormous quantity of gas is simply ejected from
the halo \it may \rm be a possibility,
although we draw attention to the fact that the most sophisticated of such
models leads inevitably to the identical halo
abundance discrepancies (\ie [C,N/O]$\simgt +0.5$) illustrated in our study.

\acknowledgments

We wish to thank both Gilles Chabrier and Tim Axelrod for a number of
helpful suggestions.  BKG
acknowledges the financial support of NSERC, through its
Postdoctoral Fellowship program.



\begin{thebibliography}{}

\bibitem[Adams \& Laughlin 1996]{AG96}
Adams, F.C. \& Laughlin, G. 1996,
\apj, 468, 586

\bibitem[Alcock \etal 1993]{A93}
Alcock, C., \etal 1993,
\nat, 365, 621

\bibitem[Alcock \etal 1997]{A97}
Alcock, C., \etal 1997,
\apj, in press

\bibitem[Arnett 1996]{Dave96}
Arnett, D. 1996,
Formation of the Galactic Halo, ed. H. Morrison \& A. Sarajedini, ASP Conf.
Series, 337

\bibitem[Chabrier, Segretain \& M\'era 1996]{CSM96}
Chabrier, G., Segretain, L. \& M\'era, D. 1996,
\apj, in press

\bibitem[Charlot \& Silk 1995]{CS95}
Charlot, S. \& Silk, J. 1995,
\apj, 445, 124

\bibitem[Dearborn \etal 1986]{DLADHMG86}
Dearborn, D.S.P., Liebert, J., Aaronson, M., Dahn, C.C., Harrington, R., Mould,
J. \& Greenstein, J.L. 1986,
\apj, 300, 314

\bibitem[De Kool \& Green 1995]{DG95}
De Kool, M. \& Green, P. 1995,
\apj, 449, 235

\bibitem[De Paolis \etal 1997]{DP97}
De Paolis, F., Ingrosso, G., Jetzer, Ph. \& Roncadelli, M. 1997,
Dark and Visible Matter in Galaxies and Cosmological Implications, ed. M.
Persic \& P. Salucci, ASP Conf. Series, in press

\bibitem[Eggen, Lynden-Bell \& Sandage 1962]{ELS62}
Eggen, O., Lynden-Bell, D. \& Sandage, A. 1962,
\apj, 136, 748

\bibitem[Fields, Mathews \& Schramm 1996]{FMS96}
Fields, B.D., Mathews, G.J. \& Schramm, D.N. 1996,
\apj, in press

\bibitem[Freeman 1996]{F96}
Freeman, K.C. 1996,
Formation of the Galactic Halo, ed. H. Morrison \& A. Sarajedini, ASP Conf.
Series, 3

\bibitem[Gibson 1996a]{G96a}
Gibson, B.K. 1996a,
\mnras, 278, 829

\bibitem[Gibson 1996b]{G96b}
Gibson, B.K. 1996b,
\apj, 468, 167



\bibitem[Gratton \& Caretta 1996]{GC96}
Gratton, R. \& Caretta, E. 1996,
Formation of the Galactic Halo, ed. H. Morrison \& A. Sarajedini, ASP Conf.
Series, 371

\bibitem[Green \& Margon 1994]{GM94}
Green, P. \& Margon, B. 1994,
\apj, 423, 723

\bibitem[Greggio \& Renzini 1983]{GR83}
Greggio, L. \& Renzini, A. 1983,
\aap, 118, 217

\bibitem[Hegyi \& Olive 1986]{HO86}
Hegyi, D.J. \& Olive, K.A. 1986,
\apj, 303, 56

\bibitem[Fujimoto \etal 1995]{FSIH95}
Fujimoto, M.Y., Sugiyama, K., Iben, Jr., I. \& Hollowell, D. 1995,
\apj, 444, 175

\bibitem[Langer \& Henkel 1995]{LH95}
Langer, N. \& Henkel, C. 1995,
\ssr, 74, 343

\bibitem[Larson 1986]{L86}
Larson, R.B. 1986,
\mnras, 218, 409

\bibitem[Maeder 1992]{M92}
Maeder, A. 1992,
\aap, 264, 105

\bibitem[Marigo, Bressan \& Chiosi 1996]{MBC96}
Marigo, P., Bressan, A. \& Chiosi, C. 1996,
\aap, in press

\bibitem[M\'era, Chabrier \& Schaeffer 1996]{MCS96}
M\'era, D., Chabrier, G. \& Schaeffer, R. 1996,
Europhys. Lett., 33, 327

\bibitem[Renzini \& Voli 1981]{RV81}
Renzini, A. \& Voli, M. 1981,
\aap, 94, 175

\bibitem[Ryu, Olive \& Silk]{ROS90}
Ryu, D., Olive, K.A. \& Silk, J. 1990,
\apj, 353, 81

\bibitem[Salpeter 1955]{S55}
Salpeter, E.E. 1955,
\apj, 121, 161

\bibitem[Schaller \etal 1992]{SSMM92}
Schaller, G., Schaerer, D., Meynet, G. \& Maeder, A. 1992,
\aaps, 96, 269

\bibitem[Timmes, Woosley \& Weaver 1995]{TWW95}
Timmes, F.X., Woosley, S.E. \& Weaver, T.A. 1995,
\apjs, 98, 617

\bibitem[Tornamb\`e 1989]{T89}
Tornamb\`e, A. 1989,
\mnras, 239, 771

\bibitem[Turatto, Cappellaro \& Benetti 1994]{TCB94}
Turatto, M., Cappellaro, E. \& Benetti, S. 1994,
\aj, 108, 202

\bibitem[Van Buren 1985]{V85}
Van Buren, D. 1985,
\apj, 294, 567

\bibitem[van den Hoek \& Groenewegen 1996]{VG96}
van den Hoek, L.B. \& Groenewegen, M.A.T. 1996,
\aap, in press

\bibitem[Woosley \& Weaver 1995]{WW95}
Woosley, S.E. \& Weaver, T.A. 1995,
\apjs, 101, 181

\end{thebibliography}
\end{document}